\documentclass[a4paper,12pt]{article}

\usepackage{amssymb,latexsym} 

\usepackage[utf8]{inputenc}

\usepackage{fullpage}

\usepackage{boxedminipage}

\usepackage{listings}

\usepackage{minitoc}

\makeatletter \@addtoreset{equation}{section} 
\makeatother

\newif\ifpdf \ifx\pdfoutput\undefined \pdffalse 

\else \pdfoutput=1 

\pdftrue \fi

\ifpdf 
\usepackage[pdftex]{graphicx} \else 
\usepackage{graphicx} \fi 
\begin{document}

\ifpdf \DeclareGraphicsExtensions{.pdf, .jpg, .tif} \else \DeclareGraphicsExtensions{.eps, .jpg} \fi 
\begin{titlepage}
	
	\thispagestyle{empty} 
	\begin{flushright}
		\hfill{CERN-PH-TH/2006-008} \\
		\hfill{hep-th/0601138} 
	\end{flushright}
	
	\vspace{35pt} 
	\begin{center}
		{ \LARGE{\bf Updates in local supersymmmetry and \\ its spontaneous breaking }}
		
		\vspace{60pt}
		
		{\bf Gianguido Dall'Agata$^\ddag$ \ and \ Sergio Ferrara$^{\ddag\star}$}
		
		\vspace{30pt}
		
		{\it Physics Department,\\
		Theory Unit, CERN, \\
		CH 1211, Geneva 23, \\
		Switzerland}
		
		\vspace{15pt}
			
		{\it INFN,\\
		Laboratori Nazionali di Frascati, \\
		Italy}

		\vspace{40pt}
		
		{ABSTRACT} 
	\end{center}
	
	\vspace{10pt} 
	
	We give a basic review of some recent developments in local supersymmetry breaking in 4-dimensional effective theories coming from compactifications of string and M-theory in the presence of non-trivial form and geometrical fluxes.
	
	\vspace{150pt} 
	
	\noindent \textit{Contribution to the Proceedings of the 43rd Erice International School of Subnuclear Physics ``Towards New Milestones in our Quest to go Beyond the Standard Model'' (29 August--7 September 2005)}
\end{titlepage}
\newpage \baselineskip 6 mm

\tableofcontents

\section{Introduction}

One of the main properties of spontaneously broken supersymmetric field theories is that the scalar potential which determines the classical spectrum of the theory is, to a large extent, controlled by supersymmetric Ward-identities \cite{Ferrara:1985gj,Cecotti:1984wn,D'Auria:2001kv}.
More importantly, for theories with $N \geq 2$ supersymmetry the low-energy scalar potential is completely fixed by the gauging \cite{D'Auria:2001kv} of some invariance group of the theory up to certain ``integration constants'' usually called ``Fayet--Iliopoulos terms'' \cite{Fayet:1974jb}.
In $D = 4$, an effective action from superstring or M-theory compactifications can be described in full generality if a certain number of local supersymmetries are preserved in the process of compactification \cite{Zumino:1979et}.
In the following paragraphs we will consider in some detail situations where this kind of possibility is realized.
If $N$-extended local supersymmetries are preserved, the classical vacua of the theory are then selected by the scalar potential.
Supergravity relates such potential to the mass deformations of the theory.
The latter are in turn connected to the gauging procedure, if the theory has at least two unbroken supersymmetries.

Spontaneously broken theories of local supersymmetry have recently received a lot of attention, because some new compactifications of superstring and M-theory can be described in a rather general way and some of the new vacua are obtained by turning on different types of fluxes of some geometrical objects of the higher-dimensional theory, such as form fluxes or fluxes of the spin connection along the directions of the ``internal manifold'' of compactification.

The new geometrical data of the theory can then be related to the massive deformations of the lower-dimensional supergravity and it is often possible to derive general conclusions on the mass spectrum and on the broken and unbroken symmetries of the new vacua.

This review is organized as follows.
In the first part (sections 2,3 and 4), we recall the so-called flow equations.
These are essentially the content of the supersymmetric Ward-identities of the theory \cite{Ferrara:1985gj,Cecotti:1984wn,D'Auria:2001kv}.
The second part (sections 5 and 6) deals with the construction of the $D=4$ supergravity action with the geometrical data of the compactification procedure, in particular of the flux compactifications, which can be described by manifolds with ``exceptional'' $G$-structure \cite{Gauntlett:2002sc}, as an extension of manifolds with exceptional holonomy.
The most famous examples of the latter are Calabi--Yau and $G_2$-holonomy compactifications with $N=1$ or $N=2$ residual supersymmetry in four dimensions.
Finally, we briefly describe the black-hole attractor mechanism, which also falls in this class, as it is obtained by a flux potential.
The latter is generated by the wrapping of a form (5-form in type IIB) on  a product of a 2-sphere with a homology cycle of the internal manifold (a 3-cycle for Calabi--Yau 3-folds in type IIB).

\section{Mass deformations of extended supergravity}\label{sec:Massdef}

The mass terms of a 4-dimensional extended supergravity lagrangian are related to the ``Yukawa'' couplings of the spin 3/2 spin 1/2 sector and to the scalar potential of the theory.
They can be schematically written as \cite{D'Auria:2001kv} 
\begin{equation}
	{\cal L}_{m} ({\rm det}V)^{-1} = S_{AB} \bar \psi_{\mu}^A \sigma^{\mu\nu}\psi_{\nu}^B + i \, N_{I}^A \bar \lambda^{I} \gamma^{\mu}\psi_{\mu A} + M^{IJ} \bar \lambda_{I} \lambda_J + {\rm h.c.}
	- V(q).
	\label{Lmatter} 
\end{equation}

Here $S_{AB}$, $N_I^A$, $M^{IJ}$ are the fermionic ``mass matrices'', which are related to the fermionic (supergravity) transformation laws as follows: 
\begin{eqnarray}
	\delta \psi_{\mu A} &=& \ldots
	+\frac12 \, \gamma_\mu \, S_{AB} \epsilon^B\,,\\
	\delta\lambda_I &=& \ldots
	+ \, N_I^A \epsilon_A\,
\end{eqnarray}
and moreover, the scalar potential $V(q)$ is given by \cite{Ferrara:1985gj,Cecotti:1984wn} 
\begin{equation}
	\delta_B^A V(q) = -3 \bar S^{AC} S_{CB} + N_I^A N_B^I, \label{Ward} 
\end{equation}
where $(N_I^A)^{*} = N_A^I$, $S_{AB} = S_{BA} = (S^{AB})^{*}$.
If the full R-symmetry U($N$) of $N$-extended supergravity is manifest, then the index $A$ labels the $N$-dimensional representation of U$(N)$ and $\epsilon_A(x)$ is the local supersymmetry parameter of the transformations (as left-handed or Weyl spinor).
Its complex conjugate $\epsilon^{A}(x)$ is a right-handed spinor.
$N$-extended supergravity can be formulated in such a way that the R-symmetry $H$ is a ``local symmetry of the theory''; $H$ is, in fact, part of the holonomy group of the $\sigma$-model geometry of the scalar manifold of the spin 0 components of the supersymmetric multiplets of the theory.
For example in the $N=1$ theory, U(1) is part of the U($n$) holonomy of the Hodge--K\"ahler manifold \cite{Zumino:1979et,Witten:1982hu} of the $n$ chiral (Wess--Zumino) multiplets.
In $N =2$, U($2$) = U($1$) $\times$ SU($2$) is split in the U(1) part of the holonomy of the special geometry and the SU(2) part of the holonomy of the quaternionic geometry of the hypermultiplets \cite{Bagger:1983tt,Strominger:1990pd,D'Auria:1990fj}.
Moreover, in local $N=1,2$ supergravity, there are associated U(1) and SU(2) bundles \cite{Witten:1982hu,Andrianopoli:1996cm} whose curvatures are in the same cohomology class as the K\"ahler form and as triplets of quaternionic almost complex structures of the base manifold.
In simple terms, this means that in the supergravity transformations of the gravitino 
\begin{equation}
	\delta\psi_{\mu A} = D_{\mu} \epsilon_{A} + \ldots  = \partial_\mu \epsilon_A + \frac14 \omega_\mu^{ab} \gamma_{ab} \epsilon_A + \omega_{\mu A}{}^B \epsilon_B +\ldots, \label{gravsusy} 
\end{equation}
the covariant derivative contains a term $\omega_{\mu A}{}^{B}$ that is the U(N) connection of the principal U(N) bundle.
For symmetric spaces the U(N) connection is identified with the $H$ connection of the $G/H$ coset space.
For $N=1$ and $N=2$ this is the U(1) and U(2) connection of Hodge--K\"ahler and quaternionic geometry, respectively.
All fermions of the theory are assigned to representations of $H$.
Local $H$ symmetry then determines their couplings to the non-linear $\sigma$-model of the scalar manifold.
For rigid supersymmetry, the $H$ connection trivializes (becomes flat) \cite{Andrianopoli:1996cm}.
In such a case the Hodge--K\"ahler structure becomes simply K\"ahler and the quaternionic structure becomes hyper-K\"ahler.

For all $N>2$ theories, the $\sigma$-model geometry is trivial (flat) in the rigid case and becomes (at least locally) a symmetric space in the case of local supersymmetry (supergravity).
The local R-symmetry $H$ becomes part of the isotropy group of the symmetric space $G/H^{\prime}$, where $H^{\prime} = H \times H_{\rm Matter}$.
Here $H_{\rm Matter}$ is the part of the isotropy group related to matter multiplets, which can only exist for $N=3,4$.
In such a case $H_{\rm Matter} = $SU$(n)$ for $N=3$ and $H_{\rm Matter} = SO(n)$ for $N=4$, where $n$ is the total number of matter (vector) multiplets.

An interesting case is the $N=8$ theory.
In the local $SU(8)$ covariant formulation of $N = 8$ supergravity, the rigid symmetry of the equations of motion is $E_{7(7)}$ but the maximal rigid symmetry of the lagrangian is SL$(8, {\mathbb R})$ \cite{Hull:1984wa,Cordaro:1998tx}.
In different (duality related) formulations, the maximal rigid symmetry of the lagrangian may be a different (non-compact) subgroup of $E_{7(7)}$ and the manifest local symmetry may be a smaller subgroup of SU$(8)$.
For example, by reducing from five to four dimensions \cite{Cremmer:1979uq} the non-compact symmetry of the lagrangian is \cite{Cremmer:1979uq,Andrianopoli:2002mf} $E_{6(6)}\otimes $SO$(1,1) \rtimes T_{27}$ and the manifest local symmetry is USp(8).
By performing a Scherk--Schwarz reduction from 11 to 4 \cite{Scherk:1979zr} (and dualizing the antisymmetric tensors) the maximal non-compact symmetry of the action is \cite{Dall'Agata:2005ff,Andrianopoli:2005jv,D'Auria:2005er} $GL(7) \rtimes N_{42}$, where $N_{42}$ is a nilpotent algebra of dimension 42.
In this case the manifest local symmetry is Spin$(7)$.
In this formulation the spin 3/2 spectrum transforms as the 8 of SU(8), USp(8) and Spin$(7)$ respectively, but the covariant derivative still has the full SU(8) connection.

\section{Electric-magnetic duality}\label{sec:EMrelations}

The geometrical data of the $\sigma$-models of the scalar degrees of freedom are encoded in the K\"ahler (or special K\"ahler) geometry for $N=1$ and $N=2$ theories.
Moreover, for $N\geq 2$ theories, where scalar and vectors lie in the same multiplet, an additional property must hold, since the non-linear $\sigma$-model must be consistently coupled to the vector fields and to the fermions.
This requirement demands the existence \cite{Gaillard:1981rj,Andrianopoli:1996ve} in the theory of a ``flat symplectic bundle'' Sp$(2 n_V, {\mathbb R})$ where $n_V$ is the number of vectors in the theory, and the base manifold is the scalar manifold of which the vectors are supersymmetric partners.
This state of affairs implies that the $N=2$ Hodge--K\"ahler geometry of vector multiplets is ``special K\"ahler'' \cite{deWit:1984px,Andrianopoli:1996cm}.
For $N > 2$ the symmetric spaces $G/H^{\prime}$ of extended supergravity are of a very special type.
In particular the group $G$ must have a ``symplectic representation'' $R_{V}$ to which the vector field strengths ${\cal F}^{\Lambda}$ and their dual ${\cal G}_{\Lambda}$ belong.
${\cal G}_{\Lambda}$ is defined \cite{Gaillard:1981rj} through the geometrical equation ${\cal G}_{\Lambda} = 2 \frac{\delta {\cal L}}{\delta {\cal F}^{\Lambda}}$, where the lagrangian ${\cal L} = {\cal L}({\cal F}^{\Lambda}, \chi, 
\partial \chi)$ is function of the field strength ${\cal F}^{\Lambda}$ and of other fields (spin 0, 1/2, 3/2) in the theory.

Since the representation $R_V$ is symplectic, the coset representative $L(q)$ is a symplectic matrix 
\begin{equation}
	L = \left(
	\begin{array}{cc}
		{\cal U} & {\cal V} \\
		{\cal W} &{\cal Z}
	\end{array}
	\right)\,, \quad 
	\begin{array}{l}
		{\cal U}^{T} {\cal W}, {\cal V}^T {\cal Z} \quad {\rm symmetric} \\
		{\cal U}^{T} {\cal Z} -{\cal W}^T{\cal V} = 1.
	\end{array}
\end{equation}
where ${\cal U,V,W,Z}$ are $n_v \times n_v$ real matrices.
For lagrangians quadratic in the field strength, the normalization of the ``kinetic terms'' of the vector fields is given through the formula 
\begin{equation}
	{\cal L} = {\cal F}^{\Lambda} \wedge {\cal G}_{\Lambda}, \label{Lr} 
\end{equation}
with ${\cal G}_{\Lambda} = {\rm Re}\; {\cal N}_{\Lambda\Sigma}\; {\cal F}^{\Sigma} + \frac12 {\rm Im}\; {\cal N}_{\Lambda \Sigma}\;{\cal F}^{-\Sigma}$ or, in complex notation :
\begin{equation}
	{\cal L} = {\rm Im}\; {\cal F}^{-\Lambda} \overline{\cal N}_{\Lambda \Sigma}{\cal F}^{-\Sigma} = {\rm Im}\; {\cal F}^{-\Lambda} \wedge {\cal G}^-_{\Lambda}.
	\label{Lc} 
\end{equation}
The complex symmetric matrix ${\cal N}_{\Lambda \Sigma} = {\cal N}_{\Sigma\Lambda}$ with ${\rm Im}\; {\cal N} < 0$ is related to the coset representative through \cite{Gaillard:1981rj,Andrianopoli:1996ve} 
\begin{equation}
	{\cal N} = ({\cal W} -i{\cal Z})({\cal U} - i{\cal V})^{-1} = h f^{-1}, \label{boh} 
\end{equation}
and it is subject to a fractional transformation 
\begin{equation}
	{\cal N}^{\prime} = (C+ D {\cal N})(A + B {\cal N})^{-1} \label{boh2} 
\end{equation}
under a symplectic transformation of parameters $A,B,C,D$ ($A^{T} C$, $B^{T}D$ symmetric and $A^{T} D - C^TB = 1$) under which the vector field strengths rotate 
\begin{equation}
	\left(
	\begin{array}{c}
		{\cal F}^{\prime} \\
		{\cal G}^{\prime}
	\end{array}
	\right) = \left(
	\begin{array}{cc}
		A &B \\
		C & D
	\end{array}
	\right)\left(
	\begin{array}{c}
		{\cal F} \\
		{\cal G}
	\end{array}
	\right).
	\label{dual} 
\end{equation}
Here $S =  {\scriptsize \left(
\begin{array}{cc}
	A &B \\
	C & D
\end{array}
\right)}$ satisfies $S^{T} \Omega S = \Omega$, where $\Omega = {\scriptsize \left(
\begin{array}{cc}
	0 &1 \\
	-1 & 0
\end{array}
\right)}$.
For infinitesimal (continuous) transformations where $S = 1 + s$, with $s = {\scriptsize \left(
\begin{array}{cc}
	a &b \\
	c & -a^T
\end{array}
\right)}$, $b = b^{T}$, $c= c^T$, and $a$ arbitrary, ${\cal N}$ transforms as 
\begin{equation}
	\delta {\cal N} = - a^{T} {\cal N} - {\cal N} a + c - {\cal N} b {\cal N}.
	\label{transf} 
\end{equation}

We observe that under an $S$ transformation the lagrangian ${\cal L}$ becomes
\begin{equation}
	\delta {\cal L} = \frac14 \left({\cal F} c \widetilde {\cal F} + {\cal G} b \widetilde {\cal G}\right).
	\label{transaction} 
\end{equation}
If $b=c=0$, ${\cal L}$ is invariant; if $b = 0$, $c \neq 0$ ${\cal L}$ is invariant up to a total derivative, while if $b \neq 0$, ${\cal L}$ is not invariant.
In the absence of other fields, such as antisymmetric tensors, the requirement for the gauging is that $b = 0$.
If $c \neq 0$ and it is a local symmetry, i.e.~$c = c(x)$, the invariance can be restored up to a total derivative, provided certain conditions, discussed in the next sections, are fulfilled.
If the Sp$(2 n_v, {\mathbb R})$ is an element of $G$ then ${\cal N}^{\prime}(\phi^{\prime})$ since in this case the duality rotation is an isometry of the non-linear $\sigma$-model.
For example, in $N=8$, $D=4$ supergravity, with $G = E_{7(7)}$, the Sp$(56,{\mathbb R})$ duality group contain $G$ according to the embedding $56 \to 56$.
For a coordinate transformation of $G/H$ $\phi \to \phi^{\prime}$, which belongs to $G$,
\begin{equation}
	\delta q^{u} = \xi^{A} k_A^u (q), \label{delta} 
\end{equation}
where $k_A^u(q)$ are the ``Killing vectors'' of $G$, $\xi^A = \xi^A(a,b,c)$.
The group $G$ has no linear action on the gauge potentials $A^{\Lambda}_{\mu}$ unless $b=0$, in which case 
\begin{equation}
	\begin{array}{rcl}
		\delta A_{\mu}^{\Lambda} &=& a^\Lambda{}_{\Sigma} A^{\Sigma}_{\mu},\\[2mm]
		\delta {\cal N} &=& c - a^{T} {\cal N} - {\cal N}a .
	\end{array}
	\label{varie} 
\end{equation}

\section{The gauging of duality rotations}\label{sec:gauging}

For a duality rotation to be gauged, in a theory described only in terms of scalar and vector fields as bosonic degrees of freedom, the duality rotations must be of the lower triangular form ($b=0$).
Moreover the matrices $a_{\Sigma}^{\Lambda}(x)$, $c_{\Lambda\Sigma}(x)$ must fulfil  additional restrictions, which come from the gauging, namely:
\begin{equation}
	\begin{array}{rcl}
		a^{\Lambda}_{\Sigma}(x) &=& f^{\Lambda}_{\Sigma\Delta}\xi^{\Delta}(x),\\
		c_{\Lambda\Sigma}(x) &=& c_{\Lambda\Sigma,\Delta}\xi^{\Delta}(x), 
	\end{array}
	\label{agau} 
\end{equation}
where $f^{\Lambda}_{\Sigma\Delta}$ are the structure constants of the gauge group ${\cal G} \subset G$ and $c_{\Lambda\Sigma,\Delta}$ are ``constants'' that must satisfy the following properties \cite{deWit:1984px} 
\begin{eqnarray}
	c_{(\Lambda\Sigma,\Delta)} &=& 0, \label{prop1}\\
	f_{\Gamma(\Pi}^\Delta c_{\Lambda)\Delta,\Sigma} - f_{\Sigma(\Pi}^\Delta c_{\Lambda)\Delta,\Gamma} + \frac12 f_{\Gamma\Sigma}^\Delta c_{\Pi\Lambda,\Delta} &=& 0.
	\label{prop2} 
\end{eqnarray}
The last equation demands that $c$ be a non-trivial cocycle of the Lie algebra of ${\cal G}$.
The first equation has to be fulfilled even if ${\cal G}$ is abelian, in which case $f^{\Lambda}_{\Sigma\Delta} =0$.
If $c_{\Lambda\Sigma,\Delta} \neq 0$, translation isometries of the $\theta$ term ($\theta F \widetilde F$) are gauged, which demands the presence of Wess--Zumino terms in the action (generalized Chern--Simons terms) widely discussed in the literature \cite{deWit:1984px,Andrianopoli:2002mf,Andrianopoli:2002vy,Andrianopoli:2004sv,deWit:2005ub}.
The complex $n_V \times n_V$ square matrix $(f,h)$ determines most of the other couplings in the lagrangian and transformation laws.
For generic $N$-extended theories $f^{\Lambda} = (f^{\Lambda}_{[AB]},f^{\Lambda}_{I})$, $h_{\Lambda} = (h_{\Lambda,[AB]},h_{\Lambda,I})$, where the $f^\Lambda$ appear in the supersymmetric variation of the gauge vector potentials \cite{D'Auria:2001kv} 
\begin{equation}
	\delta A_\mu^\Lambda = 2 f^\Lambda_{[AB]}\bar \psi^A_{\mu} \epsilon^B + i f_I^{\Lambda} \bar \lambda^{IA} \gamma_\mu \epsilon_A + {\rm h.c.}
	\label{varA} 
\end{equation}
and the symplectic invariant combination 
\begin{equation}
	h^T {\cal F}^{-}_{\mu\nu} - f^T {\cal G}^{-}_{\mu\nu} \label{label} 
\end{equation}
enters the supersymmetry transformation laws of the fermions \cite{Strominger:1990pd,Ceresole:1995ca,D'Auria:2001kv}.
For $N = 2$ special geometry \cite{deWit:1984px,Andrianopoli:1996cm}, where the scalar manifold is an arbitrary special K\"ahler manifold, these matrices are \cite{Ceresole:1995ca,Ceresole:1995jg,Andrianopoli:1996cm} $f_I^{\Lambda} = (\bar f_{\bar \imath}^\Lambda,  L^{\Lambda})$, $h_{I\Lambda} = (\bar f_{\bar \imath\Lambda},  M_{\Lambda})$, $\Lambda = 0, \ldots, n_V$ in terms of the symplectic sector ($L^{\Lambda}, M_\Lambda$) and their covariant derivatives ($f_{i}^{\Lambda} = D_i L^\Lambda, h_{i\Lambda} = D_i M_\Lambda$).

We conclude this section by giving the scalar potential in terms of the geometrical data of the gauged supergravity.
For $N = 1$ theory the gravitino mass $S_{AB} = L$ is given by 
\begin{equation}
	L = W(z) {\rm e}^{\frac12 K(z,\bar z)}, \label{superpotN1} 
\end{equation}
and the fermionic shifts $N_{I}^{A}$ are $\left\{N^i = 2 g^{i\bar \jmath} \nabla_{\bar \jmath} \bar L, D^{\Lambda} = 2 ({\rm Im}\; {\cal N}_{\Lambda\Sigma})^{-1} P_\Sigma\right\}$.

The potential then is \cite{Cremmer:1982en}
\begin{equation}
	V = 4\left(-3 L \bar L + g^{i\bar \jmath}\nabla_i L \nabla_{\bar \jmath}\bar L - \frac{1}{16} {\rm Im}\; {\cal N}_{\Lambda\Sigma} D^\Lambda D^\Sigma\right), \label{potN1} 
\end{equation}
where $P_\Lambda$ is the prepotential of the Killing vector $k^{i}_\Lambda = i g^{i\bar \jmath}
\partial_{\bar \jmath} P_\Lambda$.
The Fayet--Iliopoulos terms correspond to a possible shift $P_{\Lambda} \to P_{\Lambda} + \xi_{\Lambda}$ for abelian factors of the gauge group ${\cal G}$.

For $N =2$ theories the geometrical data refer to both the special K\"ahler manifold of vector multiplets ($m_V$ their number; note that $n_V = m_v +1$) with metric $g_{i\bar \jmath}(z,\bar z)$ and Killing vectors $k^i_{\Lambda}(z,\bar z)$ and to the quaternionic manifold of hypermultiplets with metric $h_{uv}(q)$ and Killing vectors $k_{\Lambda}^u(q)$.
Because of the special nature of these manifolds, there are additional geometrical quantities that enter the scalar potential, namely the matrix $(f,h)$ defined before (for special geometry) and the three-holomorphic prepotentials \cite{D'Auria:1990fj,Andrianopoli:1996cm} $P^{x}_\Lambda$ ($x=1,2,3$) for the quaternionic manifold, defined through the relation \cite{D'Auria:1990fj,Andrianopoli:1996cm,Ceresole:2001wi} 
\begin{eqnarray}
	2 k^{u}_{\Lambda} \Omega^x_{uv} &=& \nabla_{v} P^x_{\Lambda}(q), \\
	P^x_{\Lambda}(q) &=& \frac{1}{n_H} D_u K_{\Lambda v} \Omega^{x uv},
	\label{prep} 
\end{eqnarray}
where $\Omega$ is the SU(2) curvature of the quaternionic manifold with holonomy group SU(2) $\times$ USp(2 $n_H$), $n_H$ being the number of hypermultiplets in the theory.

The gravitino mass $S_{AB}$ and fermionic shifts $N$ are given by \cite{Andrianopoli:1996cm} 
\begin{eqnarray}
	S_{AB} & = & i \frac12 P^x_{\Lambda} L^{\Lambda} \sigma^{x}_{AB} = i \frac12 P_{AB\Lambda} L^{\Lambda} \sigma^{x}_{AB} = i \frac12 P_{AB}, \label{label1} \\[2mm]
	W^{iAB} & = & i \nabla^i P^{AB} + \epsilon^{AB} k^i, \label{label2} \\[2mm]
	N^{A}_{\alpha} & = & 2 {\cal U}_{\alpha u}^{A} k^{u}_\Lambda \bar L^\Lambda, \label{label3} 
\end{eqnarray}
where ${\cal U}_{\alpha u}^{A}$ is the vielbein of the quaternionic manifold ($A = 1,2$, $\alpha = 1,\ldots, 2 n_H$) and $k^i$ ($i = 1,\ldots, m_V$) are the Killing vectors of the special manifold.
Finally the scalar potential, through the formulae of section 2, reads \cite{Andrianopoli:1996cm} 
\begin{equation}
	V(z,\bar z,q) = \left(g_{i\bar \jmath}k^i_\Lambda k^{\bar \jmath}_\Sigma + 4 h_{uv}k^u_\Lambda k^v_\Lambda\right) \bar L^\Lambda L^\Sigma + g^{i\bar\jmath}f_i^\Lambda f_{\bar \jmath}^\Sigma P_\Lambda^x P_\Sigma^x - 3 \bar L^\Lambda L^\Sigma P_\Lambda^x P_\Sigma^x \label{potN2} .
\end{equation}
The last term is the contribution of the gravitino variation, the first and third terms are the contribution of the gaugino variation, while the second term is the contribution of the hyperino variation.
Note that by virtue of the identity 
\begin{equation}
	g^{i\bar \jmath} f_i^\Lambda f_{\bar \jmath}^{\Sigma} = -\frac12 ({\rm Im}\; {\cal N}_{\Lambda\Sigma})^{-1} - \bar L^{\Lambda}L^{\Sigma} = U^{\Lambda\Sigma}, \label{identity} 
\end{equation}
the first and third terms can be rewritten as (also using $P_\Lambda L^\Lambda= P_\Lambda \bar L^\Lambda =0$) 
\begin{equation}
	U^{\Lambda\Sigma}\left(P_\Lambda P_\Sigma + P_\Lambda^x P_\Sigma^x\right) = |\delta \lambda_i^A|^2. \label{boh3} 
\end{equation}
Fayet--Iliopoulos terms are only possible in the $P_\Lambda^x$ prepotentials $P_\Lambda^x \to P_\Lambda^x + \xi_\Lambda^x$ in the absence of hypermultiplets.
Examples corresponding to ``flat gaugings'' in the context of $N=2$ supergravity were found in \cite{Cremmer:1984hj}, then realizing the no-scale structure previously found in the context of $N=1$ supergravity \cite{Cremmer:1983bf,Lahanas:1986uc}.

\section{Supersymmetric vacua of supergravity}

Let us now see how the gauged supergravities described above can be obtained by compactifying string or M-theory with non-trivial fluxes.

We are used to think about these compactifications as expansions around classical vacua of string or M-theory, and for this reason we will first see how the new tool of group structures on the tangent bundle can be used in order to classify and construct these vacua.
However, we will also see that, in order to obtain effective 4-dimensional theories that can be described in terms of gauged supergravities, one needs only that the compactification manifold admit some globally defined spinor, not necessarily providing a vacuum of the theory.
In this case the effective potential will be generically of the runaway type, but it can still be described by the structures given previously.

Let us then review how the group structure of the tangent bundle can be used to classify and construct supersymmetric vacua of supergravity theories (we follow in part  \cite{Dall'Agata:2004nw,Gauntlett:2005bn,Behrndt:2005vi,Grana:2005jc,Joyce}).
We will start with a review of the ordinary geometric compactifications and the role of the holonomy group for their classification.
We will then show how this gets modified in the presence of fluxes and especially how the classification in terms of holonomy groups is replaced by the more general classification in terms of structure groups.

\subsection{Geometric compactifications}

When solving the supersymmetry conditions, the minimal setup that can be assumed is to set  to zero all the fields of the theory but the metric tensor.
When only the metric is non-vanishing, all the supersymmetry transformations are trivially satisfied, with the exception of the gravitino one (at least when the theory does not contain higher-derivative terms).
The latter becomes an equation imposing constraints on the geometry of the solution.

The supersymmetry requirement imposes that there exist a spinor $\eta$ which is parallel with respect to the Levi--Civita connection 
\begin{equation}
	\delta \psi_m = \nabla_m \eta = 0\,.
	\label{eq:dpsi} 
\end{equation}
By computing the integrability of this equation one obtains a set of constraints on the solution 
\begin{equation}
	[ \nabla_m, \nabla_n]\, \eta = -\frac14 \, {R_{mn}}^{pq} \gamma_{pq} \, \eta = 0\,.
	\label{eq:integr} 
\end{equation}
Equation (\ref{eq:integr}) can be interpreted as the fact that certain combinations of the tangent space generators 
\begin{equation}
	T_{mn} \equiv \frac14 \, {R_{mn}}^{pq} \gamma_{pq} 
\end{equation}
annihilate $\eta$ as well as the fact that the curvature is constrained.
The first fact implies that the holonomy of the space is generically reduced.
For what concerns the second comment we can see explicitly what happens by further contracting (\ref{eq:integr}) with one gamma matrix: 
\begin{equation}
	\gamma^{n} \gamma^{pq} R_{mnpq}\, \eta = \gamma^{npq} R_{m[npq]}\, \eta - 2 R_{mn} \gamma^{n}\eta = 0\,.
	\label{eq:ricciflat} 
\end{equation}
By using that $ R_{m[npq]} = 0$ by construction for the Levi-Civita connection, one obtains that the solution must be Ricci-flat: $R_{mn}=0$.

Let us now apply this procedure to M-theory, and look for spontaneous compactifications to 4-dimensional Minkowski space.
We therefore assume that the 11-dimensional metric is a product of 4-dimensional Minkowski space-time with a compact internal manifold: ${\cal M}_{11} = M_4 \times Y_{7}$.
According to our previous discussion, the possible $Y_7$ manifolds must be special-holonomy, Ricci-flat manifolds.
These spaces have been classified by Berger, and one can see that for obtaining minimal supersymmetry in four dimensions, one can compactify M-theory on $G_2$-manifolds, whose holonomy is contained in the group $G_2 \subset SO(7)$.
Berger's classification applies more generally to all types of solutions that can be obtained for purely geometric compactifications of any (ungauged) supergravity theory preserving some supersymmetry.

Let us now explore the geometric consequences of equation (\ref{eq:dpsi}) further.

The Levi--Civita connection appearing in (\ref{eq:dpsi}) takes values in the tangent space group Spin$(1,d-1)$ and actually, since it preserves the metric $\nabla_{M} g_{NP} = 0$, in SO$(1,d-1)$.
This is the structure group of the tangent bundle for a generic Riemannian manifold, i.e.~the group required to patch the tangent bundle over the manifold.
For the Levi--Civita connection it coincides with the holonomy group.

When solving the supersymmmetry conditions one reduces this group, as we have seen before, because the Killing spinors, solutions of (\ref{eq:dpsi}), are annihilated by some of the generators of this group.
This means that in order to patch together the tangent bundle over the manifold, only a subgroup $G \subset $SO$(1,d-1)$ is needed.
This fact is equivalent to the Killing spinor being a singlet of $G$: it does not transform under an action of its generators.
Clearly, in order to have a reduction of the structure group over the whole manifold, this invariant must be globally defined.
This is granted for the solutions of (\ref{eq:dpsi}).
The Killing spinors are parallel with respect to the Levi--Civita connection and therefore any solution of (\ref{eq:dpsi}) can be transported using this connection to any other point of the manifold (at least if this is simply connected).
This means that once the supersymmmetry conditions are solved in one patch, the solution can be extended globally over the manifold.
Moreover, since the metric is preserved by this connection, it is clear that the norm of all the invariants is preserved and the invariants whose norm is never vanishing are globally defined.

Following this discussion, any reduced group structure, and therefore any reduced holonomy group, implies the existence of a set of singlet tensor fields (or spinors) with respect to the structure group.
For instance, for the compactifications of M-theory to four dimensions, we have seen that the holonomy group of the internal manifold is reduced to $G_2$.
This means that there is one globally defined invariant spinor on the manifold, as there is only one singlet in the decomposition of the spinor representation of SO$(7)$ in terms of $G_2$ representations: ${\mathbf 8} \to \mathbf 1+ \mathbf 7$.
In the same way we can see that beside the metric tensor (that is a singlet of the general SO$(1,d)$ structure group of a Riemannian manifold), there is a 3-form field that is invariant under $G_2$: the co-associative form $\Phi$.
This can be seen again by taking the decomposition of the generic rank-3 antisymmetric tensor field with respect to $G_2$: $\mathbf{35} \to \mathbf 1 + \mathbf 7 + \mathbf{27}$.
It is useful to notice that this tensor (and its dual) can be obtained by contractions of the invariant spinor with the 7-dimensional gamma matrices: 
\begin{equation}
	\Phi_{mnp} = -i\,\eta^{\dagger} \gamma_{mnp} \eta.
	\label{phidef} 
\end{equation}
This also implies that $\Phi$ is parallel with respect to the Levi--Civita connection $\nabla \Phi =0$, by applying (\ref{eq:dpsi}).
Moreover, this condition gives the known differential conditions on the 3-form $\Phi$ to define a $G_2$-manifold: it is closed and co-closed: 
\begin{equation}
	d \Phi = 0 , \quad d \star \Phi = 0.
	\label{G2cons} 
\end{equation}

\subsection{Adding fluxes}

Adding fluxes obviously changes this situation.
A simple consequence is given by the backreaction of these fluxes onto the internal geometry.
The Einstein equation will now read 
\begin{equation}
	R_{mn} = F_{m}{}^{c_1 \ldots c_{p-1}} F_{n c_1\ldots c_{p-1}} + \ldots\,, \label{eq:Einstein} 
\end{equation}
where $F_{m_1\ldots m_p}$ is some $p$-form whose vacuum expectation value (vev) is assumed to be different from zero.
On the internal sector this generically implies that the space is no longer Ricci-flat.
But we can say more.
If we again look for supersymmetric configurations, the gravitino susy law tells us that there must exist a non-trivial spinor $\eta$ which is covariantly constant with respect to a certain connection ${\cal D}$, which now contains also the flux information.
Integrability of the gravitino supersymmetry equation now no longer implies a definite restriction of the holonomy group of the Levi--Civita connection and the manifold is not generically Ricci-flat anymore, not even with respect to the generalized connection ${\cal D}$.

However, we can still characterize the solutions in terms of the structure group of the tangent bundle, by using the properties of the new connection defined by ${\cal D}$.
For a general value of the fluxes, this connection does not lie in Spin$(1,d-1)$, as not all the terms in the gravitino susy rule can be rewritten in terms of Levi--Civita-plus-torsion terms.
Actually, it does not generically preserve the metric, defining the reduction of the structure group to SO$(1,d-1)$, 
\begin{equation}
	{\cal D}_M g_{NP} = Q_{MNP} \neq 0, \label{extra} 
\end{equation}
and the generic decomposition of the connection will contain an explicit dependence on these terms ${\cal D}_M = \nabla_M + \tau_{M}{}^{NP} \gamma_{NP} + \widetilde Q_{M}$, where $\tau$ is the contorsion tensor.
As a generic consequence the spinors that solve the supersymmetry equations are no longer globally defined.

This \textit{does not} imply that the solution does not preserve supersymmetry anymore.
In order to preserve supersymmetry one just needs to solve the supersymmetry preserving conditions on every patch of the manifold, but the solutions need not be globally non-vanishing.
A similar phenomenon appears when one is looking for solutions of the Killing vector equation on a manifold.
Consider for instance $S^2 \equiv \frac{SO(3)}{SO(2)}$.
This manifold has a local SO$(3)$ symmetry group.
This implies that in every patch one can define 3 non-vanishing vector fields that generate SO$(3)$.
At the same time, parallel transport of these fields changes their norm, as they are not parallel to the Levi--Civita connection and can therefore vanish at some point, as they actually do.
Nonetheless, the group of isometries is SO$(3)$ at each point on the manifold.
The same phenomenon takes place for the supersymmetry equations and the Killing spinors, solving the supersymmetry conditions.
In this case $N$ spinor fields $\eta$ satisfying ${\cal D}\eta =0$ define an $N$-supersymmetric background, even if some of the $\eta$ vanish at some point.
However, for the special cases where $Q = 0$, the connection ${\cal D}$ lies in Spin$(1,d-1)$, and solutions to ${\cal D}\eta = 0$ can be parallel-transported using this connection and therefore become globally defined.
In the first case, the structure group is reduced only locally, and one can use the techniques in \cite{Hull:2003mf} to classify the solutions.
In the second case, the structure group is globally reduced and the intrinsic torsion completely specifies the supersymmetric solutions.

Before discussing how this classification can be achieved, let us give an example of the previous discussion for the spheres $S^{n}$.
The spheres can be used in supergravity solutions to compactify some of the dimensions, preserving all supersymmetry.
For instance, we can compactify M-theory on $AdS_4 \times S^7$, preserving all 32 supersymmetries, or 4-dimensional supergravity to $AdS_2 \times S^2$ describing the maximally supersymmetric horizon of extremal black-hole solutions.
All the spheres admit a maximal number of Killing spinors satisfying the Killing equation: 
\begin{equation}
	{\cal D}_m \epsilon = \left(\nabla_{m} + i \Lambda\gamma_m\right)\epsilon = 0, \label{spherekill} 
\end{equation}
where $\Lambda$ is related to the radius of the sphere and, for flux compactifications, it is also related to the expectation value of the form flux.
For instance, the Fr\'eund--Rubin ansatz leading to compactifications of M-theory on a 7-sphere assumes $F_{\mu\nu\rho\sigma} = -6 \Lambda \epsilon_{\mu\nu\rho\sigma}$.
Despite this, the structure group of the sphere $S^n$ {\emph is not} generically reduced to the identity, but for the special cases of $n=3$ and $n=7$.
Let us consider first the case of $S^2$.
The structure group of the tangent bundle over the manifold is SO$(2)$ and it is not reduced by the existence of spinors solving (\ref{spherekill}).
This happens because the ${\cal D}$ connection takes values in SO$(3)$ and not simply in SO$(2)$ and therefore the norm of these spinors is not preserved when parallel-transported by ${\cal D}$.
Explicitly, the connection ${\cal D}$ can be put in the form of a standard SO$(3)$ connection ${\cal D}_m = \nabla_m + \tau_{m}{}^{np}\gamma_{np}$, for in $d=2$, $\gamma_{m} = i \epsilon_{mnp} \gamma^{np}$, with $m,n=1,2,3$ and $\gamma_3$ is the matrix defining the spinor chirality.
Also the vector fields that can be constructed from these spinors are therefore not globally defined and indeed they are precisely combinations of the Killing vectors defining the SO(3) isometry group.

The same relation between the $\gamma$-matrices applies in three dimensions.
This now implies that the connection ${\cal D}$ is always in SO(3) (i.e.~it preserves the metric) and therefore any solution to the sphere Killing equation can be extended globally.
In this case the structure group is reduced to the identity and the manifold becomes parallelizable.

Intermediate cases exist, too.
For instance $S^5$ has an $SU(2)$ structure, as there is one and only one globally defined vector on the manifold.
Still, compactifications of type IIB supergravity on $AdS_5 \times S^5$ preserves 32 supersymmetries.
The five-sphere is a Sasaki--Einstein manifold \cite{Gauntlett:2005ww}.
All these manifolds have one globally defined vector.

\medskip

A further example of what happens by turning on fluxes is given by string compactifications to 4 dimensions.
When the internal manifold is a six-dimensional space, from the existence of a globally defined spinor we can define a 2--form $J_{mn}= {\rm i} \, \eta^{\dagger} \gamma_{mn} \gamma^7 \eta$ and a 3--form $\Omega_{mnp} = {\rm i}\, \eta^{t} \gamma_{mnp} \left(1+ \gamma^7\right) \eta$.
In the case of zero fluxes we have shown that $\nabla \eta = 0$ and therefore also $\nabla J = \nabla \Omega = 0$, i.e.
$J$ and $\Omega$ are preserved by the Levi--Civita connection.
This further implies that $dJ = 0$ and $d\Omega= 0$, which means that $J$ is the K\"ahler form and $\Omega $ is the holomorphic form of a Calabi--Yau 3-fold.
When fluxes are turned on, we see that the internal manifold cannot be K\"ahler anymore, since $\nabla \eta \neq 0$ and therefore also $dJ$ will be different from zero.
The outcome is that the generic compactification in the presence of fluxes is given by a (warped) product of Minkowski space--time with a curved and non--complex internal manifold $$ {\cal M}_D = M_4 \times_{w} Y_{D-4}\,.
$$

The existence of a $G$-structure does not a priori put any constraints on the possible holonomy groups.
In particular, the failure of the holonomy of the Levi-Civita connection to reduce to $G \subset GL(n)$ is measured by the intrinsic torsion and this latter can be used to describe the $G$-structure.
Given some $G$-invariant form $\eta$ defining a $G$-structure, the derivative of $\eta$ with respect to the Levi--Civita connection, $\nabla\eta$, can be decomposed into $G$-modules.
The different types of $G$-structures are then specified by looking at which of these modules are present, if any.
One first uses the fact that there is a connection $\nabla^{(T)}$ preserving the spinor 
\begin{equation}
	\nabla^{(T)}\eta=0.
\end{equation}
Then $\nabla^{(T)}-\nabla$ is a tensor that has values in $\Lambda^1\otimes \Lambda^2$.
Since $\Lambda^2\cong $so$(d)=g\oplus g^\perp$ where $g^\perp$ is the orthogonal complement of the Lie algebra $g$ in $so(d)$, and $\eta$ is invariant with respect to $g$, we conclude that $\nabla\eta =(\nabla-\nabla^{(T)})\eta$ can be identified with an element $\tau$ of $\Lambda^1\otimes g^\perp$.
Furthermore, this element is a function only of the particular $G$-structure, independent of the choice of $\nabla^{(T)}$ and it is in one-to-one correspondence with the intrinsic torsion.
Explicitly \cite{Gauntlett:2003cy}, for a $p$-form $\eta$ 
\begin{equation}
	\nabla_m \eta_{n_1\dots n_p}= -p \;\tau_{m\,[n_1}{}^q\,\eta_{|q|n_2\dots n_p]}\,, \label{tor} 
\end{equation}
where $\tau \in\Lambda^1\otimes g^\perp$, $m$ is the 1-form index and $n$, $q$ label the 2-form $g^\perp\subset\Lambda^2$.

The search for supersymmetric solutions of string and supergravity theories demands the existence of spinors, which annihilate all the supersymmetry transformations.
In geometrical terms, such spinors are parallel with respect to a generalized connection including the Levi--Civita connection and the fluxes contributions: 
\begin{equation}
	\nabla^{(T')}\eta = 0\,.
	\label{susyreq} 
\end{equation}
This gives us the possibility of understanding whether a certain solution preserves supersymmetry or not by analysing its group structure in terms of the intrinsic torsion.
Indeed one needs its group-structure to be contained in those allowed by (\ref{susyreq}) 
\begin{equation}
	\nabla^{(T)}\subseteq\nabla^{(T')}\,.
	\label{susycon} 
\end{equation}
It is therefore very important to express supersymmetry conditions as constraints on the intrinsic torsion and at the same time to classify the possible group-structures of the candidate solutions in terms of the irreducible components of the same intrinsic torsion.
We have to stress, though, that this is still not sufficient to satisfy the equations of motion, unless one requires maximal supersymmetry.
Only in certain favorable cases one can translate the extra conditions coming from such a requirement in terms of torsion classes.

Let us see once more how this works explicitly in the case of M-theory.
The relevant components of the intrinsic torsion can be obtained by the product ${\mathbf 7} \times {\mathbf 7}$, where the first is the representation of $\Lambda^1$ and the second is the surviving representation of $g^{\perp}$.
This latter follows from the decomposition of the adjoint of $SO(7)$ under $G_2$: $\mathbf{21}\to \mathbf{14}+ {\mathbf 7}$, where $\mathbf{14}$ is the adjoint of $G_2$.
This produces four tensors $\tau \in {\mathbf 1} + {\mathbf 7}+ \mathbf{14}+ \mathbf{27}$ that specify completely the intrinsic torsion and hence the solutions.
For computational purposes it is extremely important to observe that these components are completely determined by computing the exterior differential on the invariant forms: 
\begin{eqnarray}
	d \Phi & = & X_1 \star \Phi + X_7 \wedge \Phi + X_{27}, \label{dphi} \\[2mm]
	d\star \Phi & = & \frac43 X_7 \wedge \star \Phi + X_{14} \wedge \Phi.
	\label{dstarphi} 
\end{eqnarray}
A classification of the allowed tensors $X_i$ in terms of the 4-form flux is given in \cite{Kaste:2003zd} and the general solutions of M-theory preserving some supersymmetry are discussed in this fashion in \cite{Dall'Agata:2003ir}.

\section{Data of the effective theory}

In this last part of the review we will focus the discussion on the derivation and properties of the scalar potential induced by the fluxes on the effective 4-dimensional theory.
We have seen how the backgrounds preserving some supersymmetry in the presence of fluxes can be classified and constructed, using the tool of the group structure of the tangent bundle.
Of course, we are as well interested in effective theories coming from compactifications that do not satisfy the 10- or 11-dimensional equations of motion, but still give some supercharges in the effective theory that may be spontaneously broken.
This requirement is related again to the existence of some globally defined spinors on the internal manifold (therefore implying a reduction of the group structure).

The idea is that, like the supersymmetry parameter of the previous section, all the spinor fields are reduced to effective 4-dimensional fields using these globally defined fields \cite{Becker:2002jj,Grana:2005ny}.
For instance, the transverse part of the M-theory gravitino $\Psi_\mu$ can be split in the 4-dimensional part $\psi_{\mu}$ and the internal globally defined spinors $\eta$ as $\Psi_\mu = \psi_\mu \otimes \eta + \psi_\mu^* \otimes \eta^* $.
For an $N=1$ compactification of M-theory, from the supersymmetry transformation of the 11-dimensional gravitino 
\begin{equation}
	\delta \Psi_A = \left\{ D_A[\omega] +\frac{1}{144} G_{BCDE} \left(\Gamma^{BCDE}{}_A - 8 \Gamma^{CDE} {\eta^B}_A \right) \right\} \epsilon_{11} , \label{grav.flux.var.11} 
\end{equation}
we can extract the supersymmetry transformation of the 4-dimensional field 
\begin{equation}
	\delta \psi_{\mu} = D_\mu\varepsilon_4 + \ldots + i e^{K/2} W \gamma_{\mu} \varepsilon^c_{4} ,\label{N1} 
\end{equation}
by comparison of the various terms in the reduction after integration over the internal space.
In (\ref{N1}) $\epsilon^c_4$ denotes the charge-conjugate spinor and we have emphasized only the superpotential term, neglecting in the dots the various terms with the vector fields.

The superpotential term (and in general all the shift terms) is then written as an integral over the internal space of the fluxes appearing in (\ref{grav.flux.var.11}) and the non-vanishing contractions of the gamma matrices between the globally defined spinors.
These contractions, as we saw in the previous section, describe the structure group of the internal manifold, and they are represented by globally defined forms.

\subsection{Superpotentials from $p=0,2,3,4$-form fluxes}\label{sub:supop}

The first instance of such structure is the Gukov--Vafa--Witten superpotential for Calabi--Yau + orientifolds compactifications of type IIB string theory, with 3-form fluxes turned on \cite{Gukov:1999gr,Gukov:1999ya}.
The superpotential reads 
\begin{equation}
	{\cal W} = \int G \wedge \Omega, \label{WIIB} 
\end{equation}
where the integral is performed over a Calabi--Yau 3-fold, $G=F_3 - \tau H_3$ is the sum of the Ramond--Ramond and Neveu--Schwarz 3-form, complexified with the axion/dilaton $\tau$, and $\Omega$ is the holomorphic form of the Calabi--Yau.
The K\"ahler potential describing the moduli space of the effective theory is the ordinary one for Calabi--Yau manifolds; it is the sum of the K\"ahler potential for the complex-structure deformations plus that for the K\"ahler deformations and the axion/dilaton: 
\begin{equation}
	K = K_V+ K_J + K_{\tau} = -\log -i \int \Omega \wedge \bar \Omega - \log \frac{4}{3} \int J \wedge J \wedge J - \log -i(\tau - \bar \tau).
	\label{Kahlpot} 
\end{equation}
This can clearly be uplifted to F-theory, where the superpotential will have the same form as (\ref{WIIB}), but now $G$ is a real 4-form and $\Omega$ is the holomorphic form of the 4-fold.

So far we discussed Calabi--Yau compactifications, where the moduli space is clearly defined and therefore also the K\"ahler potential of the effective theory.
The first instance of a superpotential for manifolds that are no longer Calabi--Yau was obtained for the heterotic theory compactifications with fluxes preserving $N=1$ supersymmetry in  four dimensions \cite{Cardoso:2002hd} and for half-flat compactifications of type IIA supergravity \cite{Gurrieri:2002wz}.
For the Heterotic theory, the resulting superpotential now contains also information on the deviation of the internal manifold from the Calabi--Yau condition $dJ = 0 = d \Omega$.
The expression is \cite{Cardoso:2003af,Becker:2003gq} 
\begin{equation}
	{\cal W} = \int \left(H + i dJ\right) \wedge \Omega, \label{Het} 
\end{equation}
where $H = dB + h$, $h$ being the cohomologically non-trivial flux.
It is clear that the last term in (\ref{Het}) is non-zero only for manifolds that are not Calabi--Yau, so that $dJ \neq 0$, but also generically $d \Omega \neq 0$, as can be obtained by integration by parts.

Since these papers appeared, there was a lot of study in order to understand better what  the possible contributions to the effective theory potential are from the geometric part, in addition to the fluxes.
We now have various derivations of the $N=1$ superpotentials for general compactifications of type IIA/IIB and M-theory.
These can be summarized by the following equations (see for instance \cite{Grana:2005ny,Villadoro:2005cu,Dall'Agata:2005fm,Gauntlett:2003cy}): 
\begin{eqnarray}
	{\cal W}_{IIA} &=& \int {\rm e}^{B+i J} \wedge F - \int (H+i dJ) \wedge \left(C + i{\rm e}^{-\phi}{\rm Re}\;\Omega\right) \label{WIIA} \\
	{\cal W}_{IIB} &=& \int {\rm e}^{i J} \wedge d \Omega + \int G \wedge \Omega \label{WIIB2} \\
	{\cal W}_{M} &=& \frac14 \int \left(C + i \Phi\right) \wedge \left[g + \frac12 d \left(C + i \Phi\right) \right] .\label{WMth} 
\end{eqnarray}
The IIA and IIB expressions contain the fluxes ($F = m_0+ F_2+ F_4$, $G = G_3$) and the globally defined forms defining an SU(3) structure ($J$, $\Omega$).
The M-theory superpotential contains on the other hand the $G_2$-structure $\Phi$.
It is more difficult to give an expression for the K\"ahler potential governing the geometry of the moduli spaces of these compactifications.
It is indeed less clear which are the light fields one should really consider in these compactifications and also when the truncation to these fields is consistent.
Recently some light was shed on this issue too \cite{Grana:2005ny}.
It has been argued that the K\"ahler potential for SU(3)-structure compactifications is perfectly analogous to the one for Calabi--Yau manifolds (\ref{Kahlpot}).
Analogously, the M-theory K\"ahler potential is formally the same as the one of $G_2$-holonomy compactifications.
It was explicitly obtained for $G_2$-structure manifolds, for instance in \cite{Dall'Agata:2005fm}, and it reads
\begin{equation}
	K = -3 \log \frac{1}{7} \int \Phi \wedge \star \Phi.
	\label{KMth} 
\end{equation}

In paper \cite{Grana:2005ny}, a general discussion of the effective theory of a compactification of type II supergravity on a manifold with SU(3) structure, before considering any addition of orbifolds or orientifold projections, was also presented.
This theory preserves $N=2$ supersymmetry at the level of the lagrangian, and theorefore also for this type of compactification the potential can be completely specified in terms of geometrical objects through the quaternionic prepotentials $P^x$.
These were found to be 
\begin{equation}
	\begin{array}{rcl}
		P^1 &=& \displaystyle -2{\rm e}^{\frac12 K_V + \phi_4} \int {\rm e}^{-B-iJ} \wedge d \,{\rm Re}\;\, \Omega,\\
		P^2 &=& \displaystyle -2{\rm e}^{\frac12 K_V + \phi_4} \int {\rm e}^{-B-iJ} \wedge d\, {\rm Im}\;\, \Omega,\\
		P^3 &=& \displaystyle \frac{1}{\sqrt{2}} {\rm e}^{2\phi_4} \int {\rm e}^{-B-iJ} \wedge G_{IIA} 
	\end{array}
	\label{prepIIA} 
\end{equation}
for type IIA compactifications, and 
\begin{equation}
	\begin{array}{rcl}
		P^1 &=& \displaystyle -2{\rm e}^{\frac12 K_J + \phi_4} \int \Omega \wedge d \,{\rm Re}\;\, {\rm e}^{-B-iJ},\\
		P^2 &=& \displaystyle 2{\rm e}^{\frac12 K_J + \phi_4} \int \Omega \wedge d \, {\rm Im}\; \,{\rm e}^{-B-iJ} ,\\
		P^3 &=& \displaystyle -\frac{1}{\sqrt{2}} {\rm e}^{2 \phi_4} \int \Omega \wedge G_{IIB} 
	\end{array}
	\label{prepIIB} 
\end{equation}
for IIB compactifications, where $\phi_4$ is the 4-dimensional dilaton.
This result therefore includes the special cases of $N=2$ Calabi--Yau compactifications with fluxes analysed in \cite{Polchinski:1996sm,Michelson:1996pn,Dall'Agata:2001zh,Louis:2002ny}.

\subsection{The attractor potential: superpotential from 5-form fluxes}\label{sub:attractor}

Another potential, which is also related to a holomorphic ``superpotential'', is the ($N=2$) black-hole potential (for an asymptotically flat extremal black-hole) whose extrema control the formula for the black-hole entropy/area  through the relation \cite{Ferrara:1995ih,Ferrara:1997tw} 
\begin{equation}
	S_{BH} = \frac{A_{\rm Hor}}{4} = \left.\phantom{\frac{1}{1}}\pi V_{BH}(p,q,\phi_{fix})\right|_{\frac{
	\partial V}{
	\partial \phi}=0}, \label{Sbh} 
\end{equation}
where $V$ is given in terms of the Sp$(2n,{\mathbb R})$ vector $(p^\Lambda, q_\Lambda) = {\cal P}$ and of the symplectic real matrix 
\begin{equation}
	{\cal M} = \left(
	\begin{array}{cc}
		A & B \\
		C & D
	\end{array}
	\right)\,, \label{matrix} 
\end{equation}
where 
\begin{eqnarray}
	A & = & {\rm Im}\;{\cal N} + {\rm Re}\;{\cal N} {\cal N}^{-1}{\rm Re}\;{\cal N} \\[2mm]
	B & = & -{\rm Re}\;{\cal N} {\rm Im}\;{\cal N}^{-1} \\[2mm]
	C & = & - {\rm Im}\;{\cal N}^{-1} {\rm Re}\;{\cal N} \\[2mm]
	D & = & {\rm Im}\;{\cal N}^{-1} 
\end{eqnarray}
through the formula \cite{Ceresole:1995ca,Ferrara:1996dd,Ferrara:1997tw} 
\begin{equation}
	V_{BH} = -\frac12 {\cal P}^{T} {\cal M} {\cal P}.
	\label{VBH2} 
\end{equation}
Note that ${\cal M}$ is symmetric and negative-definite whenever the complex symmetric matrix ${\cal N}$ has Im ${\cal N}<0$.
The matrices Im ${\cal N}$, Re ${\cal N}$ are the ``normalizations'' of the $F \wedge F$ and $F \wedge \widetilde F$ terms in the action, where $F^\Lambda$ are the vector field strengths.
For $N$-extended supergravity theories the potential $V$ is given by \cite{Andrianopoli:1996ve}\footnote{An exception is $N=6$ supergravity, where an additional singlet charge $Z$ appears in the supergravity multiplet other than $Z^{AB}$.} 
\begin{equation}
	V_{BH} = \frac12 |Z^{AB}|^{2} + |Z_I|^{2} ,\label{VBHex} 
\end{equation}
where $Z^{AB} = -Z^{BA}$ is the ``central-charge'' matrix of the supersymmetry algebra \cite{Haag:1974qh} 
\begin{equation}
	\left\{Q_\alpha^A, Q_\beta^B\right\} = \epsilon_{\alpha\beta} Z^{AB}(\phi) \label{susyalg} 
\end{equation}
and $Z_I$ are the ``matter'' charges of the ``vector multiplets''.
For $N=2$, $Z^{AB} = Z \epsilon^{AB}$, $Z_I = D_i Z \equiv \left(
\partial_i + \frac12 K_i\right) Z$ (recall that $\bar D_{\bar \imath} Z \equiv \left(
\partial_{\bar \imath} - \frac12 K_{\bar \imath}\right) Z= 0$) and the black-hole potential simplifies to \cite{Ferrara:1996dd,Ceresole:1995ca} 
\begin{equation}
	V_{BH}= Z \bar Z + g^{i\bar \jmath}D_i Z \bar D_{\bar \jmath}\bar Z = |Z|^2 + |D_i Z|^2 \label{potVBh} 
\end{equation}
in terms of the (covariantly holomorphic) central charge \cite{Ceresole:1995jg} 
\begin{equation}
	Z = {\rm e}^{K/2}(X^\Lambda e_{\Lambda} - F_{\Lambda}p^{\Lambda}).
	\label{Zdef} 
\end{equation}
Using the special geometry identity \cite{Ferrara:1990dp,Kallosh:2005ax} 
\begin{equation}
	{\cal P} = 2 {\rm Re}\;\left(-i \bar V Z - i D_i V g^{i\bar \jmath} D_{\bar \jmath}\bar Z\right) \label{Pdef} 
\end{equation}
in terms of the symplectic vectors \cite{Ceresole:1995jg,Ceresole:1995ca} 
\begin{equation}
	{\cal P} = \left(
	\begin{array}{c}
		m^{\Lambda}\\
		e_\Lambda
	\end{array}
	\right), \quad V = \left(
	\begin{array}{c}
		L^{\Lambda} = {\rm e}^{K/2}X^\Lambda\\
		M_\Lambda = {\rm e}^{K/2}F_\Lambda
	\end{array}
	\right) ,\label{mnah} 
\end{equation}
the ``attractor equations'' $D_i V_{BH} = 0$ can be rewritten as 
\begin{equation}
	{\cal P} = 2 {\rm Re}\;\left(-i \bar V Z + \frac{1}{2Z} D_i V D_j Z D_k Z C_{\bar \imath \bar \jmath \bar k} g^{i\bar \imath} g^{j\bar \jmath} g^{k\bar k}\right).
	\label{attractor} 
\end{equation}
For supersymmetric attractors ($D_i Z = 0$), this formula simply becomes \cite{Ferrara:1995ih,Strominger:1996kf,Ferrara:1996dd} 
\begin{equation}
	{\cal P} = 2 {\rm Re}\;\left(-i \bar V Z \right).
	\label{susyattractor} 
\end{equation}
In terms of the Calabi--Yau data $Z, D_i Z$ are given by the ``fluxes'' of the IIB 5-form along the holomorphic cycles of the 3-form cohomology \cite{Ceresole:1995ca,Moore:1998pn} 
\begin{equation}
	Z = \int_{CY \times S^2} F_5 \wedge \Omega \label{cohom} 
\end{equation}
and 
\begin{equation}
	D_i Z = \int_{CY \times S^2} F_5 \wedge D_i\Omega.
	\label{cohom2} 
\end{equation}
For supersymmetric attractors $F_5$ has only a (3,0) component.
The 10-dimensional space is the product of $AdS_2 \times S^2 \times CY_{pq}$ where $CY_{pq}$ is an ``attractor variety'' as discussed in \cite{Moore:1998pn}.
For non-supersymmetric attractors ($D_i Z \neq 0$), $F_5$ has also a (2,1) component;
these have recently received some attention  \cite{Kallosh:2005ax,Goldstein:2005hq,Tripathy:2005qp,Goldstein:2005rr,Alishahiha:2006ke}.

\bigskip

\noindent
{\bf Acknowledgments}

\medskip

\noindent
We would like to thank R.~Russo and R.~Kallosh for useful discussions.
The work of S.F.~has been supported in part by the European Community Human Potential Program under contract MRTN-CT-2004-005104 ``Constituents, fundamental forces and symmetries of the universe'', in association with INFN Frascati National Laboratories and by D.O.E.~grant DE-FG03-91ER40662, Task C.


\end{document}